\documentclass[vecphys]{svmult}		

\usepackage{amsmath}
\usepackage{makeidx}         
\usepackage{graphicx}        
\usepackage{epsfig}          
\usepackage{multicol}        
\usepackage[bottom]{footmisc}

\makeindex             
\begin{document}

\def\jcmindex#1{\index{#1}}
\def\myidxeffect#1{{\bf\large #1}}

\title*{Discrete variants of the $\phi^4$ model: exceptional discretizations, conservation laws
and related topics}
\titlerunning{Discrete variants of the $\phi^4$ model}
\author{
Sergey~V.~Dmitriev\inst{1,2}
\and
Panayotis~G.~Kevrekidis\inst{3}
}

\institute{
Institute for Metals Superplasticity Problems, Russian Academy of Sciences, Ufa, 450001 Russia \texttt{dmitriev.sergey.v@gmail.com}
\and
National Research Tomsk State University, 36 Lenin Prospekt, 634050 Tomsk, Russia
\and
Department of Mathematics and Statistics, University of
Massachusetts, Amherst, MA 01003 USA \texttt{kevrekid@math.umass.edu}
}
\maketitle
\abstract
{
Exceptional dicretizations of the $\phi^4$ model are reviewed, corresponding conservation laws are reported, and the properties of static and moving discrete kinks are discussed. Different approaches to producing such discretizations are given and unifying perspectives thereof are brought forth. It is also demonstrated that the high kink mobility in the exceptional dicretizations makes it possible to analyze kink-antikink collisions 
in the regime of high discreteness.
}

\section{Introduction}

Localized nonlinear excitations such as topological solitons and breathers play an important role in many areas of physics and very often they are considered in discrete media. For example, in  solid state physics they are used to describe domain walls, dislocations, and crowdions in crystals \cite{BookSGE,BraunKivshar,KinkSuperS,Ncrowdions,ND2018}, in macroscopic models of coupled pendulums \cite{Pendula}, in granular crystals \cite{Granula}, in the arrays of electric circuits \cite{Electric} and micromechanical cantilevers \cite{Cantilever}, among many others. In all these applications the mobility of solitary waves is an important issue, especially given that a typical discretization breaks the translational invariance of the continuum model and thus renders the discrete case far less amenable to genuine traveling dynamics.

From the physical point of view, the analysis of soliton mobility can be done in two different directions. The first is the estimation of the minimal external force needed to set a standing soliton in motion, and the second is the estimation of deceleration rate for a moving soliton in the absence of external forces.

Continuum Klein-Gordon equations such as $\phi^4$ or sine-Gordon equations support kinks that can be accelerated by any small external force, and moving kinks radiate no energy propagating at constant speed given the Lorentz invariance of such Klein-Gordon models (and of their D'Alembertian operators). This is typically not the case for the discrete Klein-Gordon equations, unless an integrable lattice is considered~\cite{Orfanidis}. For standard (generic) discretizations, discrete kinks are found to have two equilibrium configurations; the one having maximal potential energy is unstable, while the one with minimal energy is stable. The difference between kink energies in these two configurations defines the height of the so-called static Peierls-Nabarro potential (PNp) \cite{PNpot}. The maximal gradient of this potential defines the minimal force needed to accelerate a standing kink. In order to reduce the force required to accelerate a kink one has to reduce the maximal gradient of the static PNp, ideally making it zero. In a series of works by Speight and Ward, about 2 decades ago, one possible way to construct a discrete version of Klein-Gordon field with precisely zero static PNp was proposed \cite{SW,S1997,S1999}. Note that the absence of the static PNp does not ensure that a moving kink will propagate at constant velocity \cite{OxtobyPelinovsky}, which means that solution of the first mobility problem does not necessarily mean that the second problem will be solved automatically.

A moving discrete kink typically (i.e., in the case of generic discretizations) radiates small-amplitude wave packets losing its energy and being initially decelerated and eventually trapped by the lattice in a well of the PNp \cite{Peyrard}. There exist numerous
1
 attempts to analyze and reduce the radiation from a moving kink \cite{OxtobyPelinovsky,Peyrard,Zolot,BarashenkovVan,Cisneros, Aigner,Cattuto,ChampneysMal, KevrekidisWeinstein,Savin,FlachZolotaryuk, Braun,Zant,Dinda,Boesch,Ishimori,Schmidt}. The radiation is attributed to the resonance of moving kinks with small-amplitude phonons. It turns out that in the discrete Klein-Gordon systems supporting kinks, for any kink velocity there exists a phonon with the same phase velocity. Excitation of such phonons by a moving kink results in its deceleration. Nevertheless it has been shown that the radiation can vanish for a set of selected kink velocities \cite{OxtobyPelinovsky,Peyrard,Aigner,ChampneysMal,Savin,FlachZolotaryuk,Schmidt}. There exists an example of kink in a non-integrable lattice, radiating no energy while moving with an arbitrary speed \cite{BarashenkovVan}. Kink solutions with oscillating background (also known as nanopterons) have been found as permanent profile traveling waves moving with constant velocity \cite{Savin}. 

In this work we focus on the first mobility problem by describing existing approaches to discretize the Klein-Gordon field in a way such that static kinks are not trapped by the lattice and they can be accelerated by any weak external force. Following \cite{BOP2005}, such discretizations will be called {\em exceptional}. \jcmindex{\myidxeffect{E}!Exceptional discretization} The kink mobility will also be discussed, as well as the conservation laws valid for the different exceptional discretizations.

\section{Klein-Gordon field with $\phi^4$ potential and its ``standard'' discretization}
\label{KGf}

\jcmindex{\myidxeffect{D}!Discretization}

The Klein-Gordon model is described by the \jcmindex{\myidxeffect{H}!Hamiltonian} Hamiltonian
\begin{eqnarray} \label{eq:lagrangy}
H=K+E=\frac{1}{2}\int_{-\infty}^{+\infty}\phi_t^2 dx+\int_{-\infty}^{+\infty}\Big[\frac{1}{2}\phi_x^2+V(\phi)\Big]dx,
\end{eqnarray}
where the first term represents the kinetic energy, $K$, and the second one the potential energy, $E$, with $\phi(x,t)$ being the unknown scalar field of spatial and temporal coordinates $x$ and $t$, respectively. Differentiation with respect to $x$ and $t$ is denoted by the corresponding subscripts. The function $V(\phi)$ defines the on-site potential. The following equation of motion can be derived from (\ref{eq:lagrangy}) using Hamilton's equations:
\begin{eqnarray} \label{eq:EOM}
\phi_{tt}+\frac{dV}{d\phi}-\phi_{xx}=0.
\end{eqnarray}
For static problems, the equation of motion (\ref{eq:EOM}) reduces to
\begin{eqnarray} \label{eq:staticform}
\phi_{xx}=\frac{dV}{d\phi}.
\end{eqnarray}
The on-site potential of the $\phi^4$ model has the form 
\begin{eqnarray} \label{eq:Phi4Potential}
V(\phi)=\frac{1}{2}(1-\phi^2)^2,
\end{eqnarray}
which has two energetically equivalent minima at $\phi=\pm 1$ separated by the potential barrier of height 1/2.

The $\phi^4$ equation defined by (\ref{eq:EOM}) and (\ref{eq:Phi4Potential}) supports the 
\jcmindex{\myidxeffect{K}!Kink}  (anti)kink solution
\begin{eqnarray} \label{eq:Kink}
\phi(x,t)=\pm\tanh\frac{x-x_0-vt}{\sqrt{\left(1-v^2\right)}},
\end{eqnarray}
moving with any velocity $-1<v<1$ (given the Lorentz invariance) starting at $t=0$ from $x=x_0$. The kink (antikink) interpolates between two minima of the on-site potential (\ref{eq:Phi4Potential}), from -1 to 1 (from 1 to -1). Kinks and antikinks propagate radiating no energy but their collisions are inelastic since the $\phi^4$ equation is non-integrable, as is discussed in some detail in other Chapters within this special volume. Due to translational invariance of the continuum nonlinear Klein-Gordon equation, a static kink ($v=0$) can be placed at any $x=x_0$ and any small external force $f$ in the equation of motion $\phi_{tt}+dV/d\phi-\phi_{xx}=f$ will accelerate the kink.

\begin{figure}
\begin{center}
\includegraphics[width=7cm]{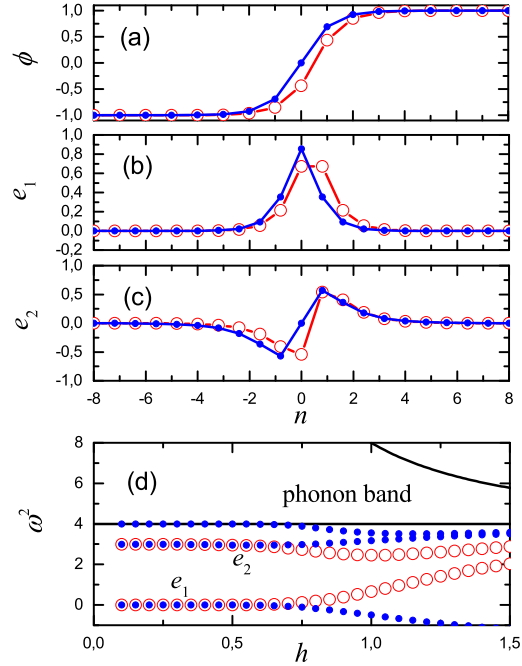}
\end{center}
\caption{Standard model defined by (\ref{eq:Classic}), (\ref{H1}): (a) Kink profile, (b) Goldstone (lowest frequency internal) mode associated with translation in the continuum limit, and (c) kink vibrational mode for $h=0.8$. (d) Frequencies of the kink's
internal modes for different magnitudes of the discreteness
parameter $h$. Two solid lines show the borders of the spectrum of
vacuum, Eq. (\ref{SpecVacClassic1}). Results for the on-site and inter-site  kinks are shown by dots and circles, respectively. The inter-site kink
is stable, while the on-site kink is unstable because the vibrational spectrum contains one
pair of imaginary frequencies.}
\label{fig1}
\end{figure}

\begin{figure}
\begin{center}
\includegraphics[width=8cm]{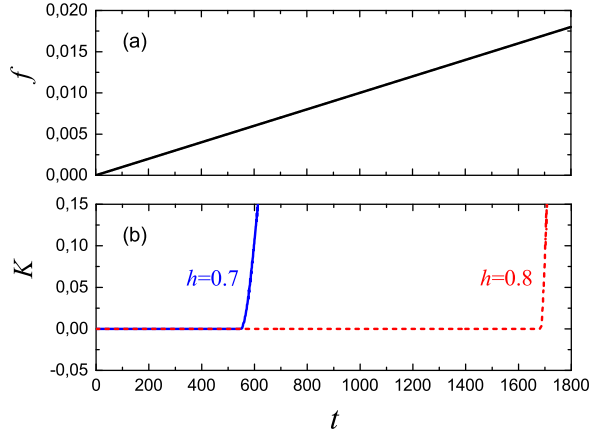}
\end{center}
\caption{Results for the classical model (\ref{eq:Classic}). (a) External force as a function of time. (b) Kink kinetic energy as the function of time. Results for $h=0.7$ (solid line) and $h=0.8$ (dashed line).}
\label{fig2}
\end{figure}

In order to discretize the Klein-Gordon equation, the lattice $x=nh$ is introduced, where $n=0,\pm1,\pm2,...$ and $h$ is the lattice spacing. The most straightforward discretization of the $\phi^4$ equation reads \cite{Campbell}
\begin{eqnarray} \label{eq:Classic}
\ddot{\phi}_n=\frac{1}{h^2}(\phi_{n-1}-2\phi_n+\phi_{n+1})-V^\prime (\phi_n) \nonumber \\
=\frac{1}{h^2}(\phi_{n-1}-2\phi_n+\phi_{n+1})+2\phi_n(1-\phi_n^2),
\end{eqnarray}
1

where $\phi_n(t)=\phi(nh,t)$ and overdot means differentiation with respect to time. 

This discretization conserves the \jcmindex{\myidxeffect{H}!Hamiltonian} Hamiltonian (total energy)
\begin{eqnarray} \label{H1}
   H_1= \frac{h}{2} \sum_n \left[ \dot{\phi}_n^2
   + \left(\frac{\phi_{n+1}-\phi_n}{h}\right)^2
   +V(\phi_n)  \right] \nonumber \\
   = \frac{h}{2} \sum_n \left[ \dot{\phi}_n^2
   + \left(\frac{\phi_{n+1}-\phi_n}{h}\right)^2
   +\left(1 - \phi_n^2 \right)^2  \right].
\end{eqnarray}
Model (\ref{eq:Classic}) does not support an exact moving kink solution. As for the static kinks, they exist only for the two symmetric configurations, centered either on a lattice site (unstable) or in the middle between two neighboring sites (stable). Such static solutions can be found numerically, e.g., by minimizing the potential energy of the solution with the use of the steepest gradient method, or by using fixed point iterations in the steady state form of Eq.~(\ref{eq:Classic}). As a first approximation, one can use the following {\em approximate} static kink solution
\begin{eqnarray} \label{phi4kinkApprox}
\phi_n= \pm \tanh [h(n - x_{0})].
\end{eqnarray}
The minimization for $x_0=0$ ($x_0=1/2$) will produce the on-site (inter-site) kink. In Fig.~\ref{fig1}(a) equilibrium kink profiles are shown for relatively high discreteness parameter of $h=0.8$. 

It is instructive to study small-amplitude vibrations of the lattice with a static kink. For this we introduce the ansatz
$\phi_n(t)=\phi_n^0+\varepsilon_n(t)$ (where $\phi_n^0$ is an
equilibrium kink solution and $\varepsilon_n(t)$ is a small
perturbation), and \jcmindex{\myidxeffect{L}!Linearization} linearize (\ref{eq:Classic}) with respect to
$\varepsilon_n$. The result is
\begin{eqnarray} \label{LinClassic1}
   \ddot{\varepsilon}_n=\Delta_2\varepsilon_{n} + 2\varepsilon_{n}
   -6 (\phi_n^0)^2 \varepsilon_{n}.
\end{eqnarray}
Here, $\Delta_2 \varepsilon_{n}=(1/h^2) (\varepsilon_{n+1}+\varepsilon_{n-1}-2 \varepsilon_n)$
stands for the discrete Laplacian.
For the small-amplitude phonon waves, $\varepsilon_{n}=\exp(i k n -i
\omega t)$, where $\omega$ is the frequency and $k$ is the wave number, from (\ref{LinClassic1}) one finds the following dispersion
relation:
\begin{eqnarray} \label{SpecClassic1}
   \omega^2=\frac{4}{h^2}\sin^2\left( \frac{k}{2} \right)
   -2 + 6(\phi_n^0)^2.
\end{eqnarray}
From (\ref{SpecClassic1}), the vacuum solution $\phi_n^0=\pm 1$ has the spectrum 
\begin{eqnarray} \label{SpecVacClassic1}
   \omega^2=4+\frac{4}{h^2}\sin^2\left( \frac{k}{2} \right).
\end{eqnarray}

We solve the eigenproblem (\ref{SpecClassic1}) numerically for the chain of $N=201$ particles 
with a static kink in the middle of the chain. The solution gives $N$ eigenfrequencies $\omega^2_n$ and $N$ eigenvectors $e_n$. Most of the eigenmodes have frequencies coinciding with the phonon spectrum of vacuum, (\ref{SpecVacClassic1}), as expected by Weyl's spectral theorem. There exist a few eigenmodes localized around the kink with frequencies below the phonon spectrum of vacuum. In Fig.~\ref{fig1}(b,c) we show the two lowest frequency eigenmodes, $e_1$ and $e_2$, localized around the kink (dots for the on-site and circles for the inter-site kink). In Fig.~\ref{fig1}(d) the upper and lower edges of the phonon spectrum of vacuum are shown by the solid lines as a function of the lattice spacing $h$. Dots and circles show the frequencies of the eigenmodes localized around the on-site and inter-site kink, respectively. The mode $e_1$ in (b) corresponds to the kink translational mode (associated with
a symmetry and a so-called Goldstone mode corresponding to zero frequency in the continuum limit) and for small $h$ it deviates exponentially weakly from the limit value, i.e., as $e^{-C/h}$~\cite{pla} for a suitable constant $C$. The lowest squared eigenfrequency of the on-site mode is negative, which becomes clear for increasing $h$ in Fig.~\ref{fig1}(d). This indicates the instability of the on-site kink configuration. For the inter-site kink, the (positive) lowest squared frequency of this mode increases with $h$ \cite{pla}. This frequency is the kink oscillation frequency in the well of the PNp. The mode $e_2$ in (c) is the well-known kink's internal mode which in the continuum $\phi^4$ equation has frequency $\omega=\sqrt 3$ \cite{Campbell,anninos,Campbell1,goodman}.

A direct calculation using the kink ansatz (cf. Eq.~(\ref{phi4kinkApprox})) inside the energy of
Eq.~(\ref{H1}) shows that the energy for the discrete $\phi^4$ kink in the \jcmindex{\myidxeffect{P}!Peierls-Nabarro potential} PNp can be approximated by
\begin{equation} \label{eq:PNpanalit}
H_0=\frac{1}{2} M\dot{X}^2+ \frac 43 +\alpha(h)\cos\left(\frac{2\pi X}{h}\right), \quad M=\frac{4}{3}, \quad \alpha(h)=\frac{8\pi^2}{3h^3}\frac{h^2+\pi^2}{\sinh(\pi^2/h)},
\end{equation}
where $X$ and $M$ are the coordinate of the center of mass and the mass of the kink, respectively. From this, one finds
\begin{equation} \label{eq:EFw}
E_0=2\alpha, \quad F_0=\alpha \frac{2\pi}{h}, \quad \omega_0=\frac{\pi}{h}\sqrt{3\alpha},
\end{equation}
which are the height of PNp, the maximal gradient of PNp, and the frequency of small-amplitude oscillations of the kink near the well of PNp, respectively. 


For $h=0.8$ from (\ref{eq:PNpanalit}) and (\ref{eq:EFw}) we have $\alpha=4.74\times 10^{-3}$, $E_0=9.48\times 10^{-3}$, $F_0=0.0372$, and $\omega_0=0.468$. Numerically we find that for $h=0.8$ the on-site (inter-site) kink has potential energy 1.3058 (1.2973). The difference gives the height of PNp of $8.5\times 10^{-3}$, which is only 10\% smaller than the estimated value. With decreasing $h$ the accuracy of the formula (\ref{eq:PNpanalit}) rapidly increases. For example, for $h=0.7$ the estimation (\ref{eq:EFw}) gives $E_0=2.39\times 10^{-3}$, while the numerically found value is $2.33\times 10^{-3}$, which is just 2.5\% smaller. Numerical evaluation of the small-amplitude vibration frequency of the kink in the well of PNp for $h=0.8$ gives the value of 0.435, which differs by 7\% from the above estimation. For $h=0.7$ the error of the frequency estimation reduces to 3.7\%.

Let us estimate numerically the minimal external force needed for a kink to overcome the PNp. To the right-hand side of equation of motion (\ref{eq:Classic}) we add the external force $f$:
\begin{equation} \label{eq:ClassicForced}
\ddot{\phi}_n=\frac{1}{h^2}(\phi_{n-1}-2\phi_n+\phi_{n+1})+2\phi_n(1-\phi_n^2)+f,
\end{equation}
assuming that the external force increases linearly with time, $f=f_0t$. To achieve a quasi-static loading we use a small value of $f_0=10^{-5}$. As initial condition, we use the equilibrium inter-site kink solution. 

In Fig.~\ref{fig2}(a) $f(t)$ is shown, while in (b) the kink kinetic energy is plotted as a function of time for the discreteness parameter $h=0.7$ (solid line) and $h=0.8$ (dashed line). While the kink is trapped in a well of the PNp, its kinetic energy is nearly zero. For $h=0.8$, the kink starts to move at $t=1690$ when the external force reaches the value of $f=0.0169$. The value $2f=0.0338$ gives the numerical estimation of the maximal gradient of PNp, which was found to be $F_0=0.0372$ from (\ref{eq:EFw}). The difference is 9\% and for $h=0.7$ the difference reduces to 3\%.
This agreement is exceptionally good, especially taking into consideration the non-stationary nature of the process. 

From the above results, we have seen that if the $\phi^4$ field is discretized as given by (\ref{eq:Classic}), a stable static kink solution exists only for the inter-site configuration and this kink is trapped by the lattice since the translational Goldstone (neutral) mode is destroyed. To set a standing kink in motion, an \jcmindex{\myidxeffect{E}!External force} external force above a threshold value can be applied; see 
also~\cite{kladko,goldst}. The effects of discreteness are rather small for $h<0.5$ and they grow in the above mentioned exponential functional form with increasing $h$. 

In the following section, the existing ways of obtaining exceptional discretizations with kinks not trapped by the lattice will be described.

\section{Approaches to derive discrete $\phi^4$ models free of the static Peierls-Nabarro potential}

There are two main approaches to the derivation of exceptional discretizations of the Klein-Gordon field equation. The first concerns the discretization of the Hamiltonian (\ref{eq:lagrangy}) and derivation of the corresponding discrete equations of motion with the use of the well-known (Hamilton equation based) technique. In this case the energy-conserving models are obtained. The alternative way involves the discretization of the equation of motion (\ref{eq:EOM}) directly. The resulting discrete equations can be Hamiltonian or non-Hamiltonian (not conserving energy). On the other hand, such discrete models can be constructed so as to conserve \jcmindex{\myidxeffect{M}!Momentum} momentum in the form \cite{K2003}
\begin{eqnarray} \label{mom4}
P=\sum_n \dot{\phi}_n (\phi_{n+1}-\phi_{n-1}),
\end{eqnarray}
or in a different form.

The standard discretization of the $\phi^4$ equation in the form of (\ref{eq:Classic}) employs a local discretization of the on-site potential (i.e., only the $n$-th site is involved). In most of the exceptional models, $V(\phi)$ or $V^\prime(\phi)$ will be discretized in a way involving three neighboring sites, so that the general form of the discretization will be
\begin{equation}
\ddot{\phi}_n=\frac{1}{h^2}(\phi_{n-1}-2\phi_n+\phi_{n+1})+F(\phi_{n-1},\phi_n,\phi_{n+1}).
\label{eq:ThreePoint}
\end{equation}
However, there will be an example of a Hamiltonian exceptional discretization with $F(\phi_{n-1},\phi_n,\phi_{n+1})=F(\phi_n)$.

\subsection{\jcmindex{\myidxeffect{S}!Speight-Ward Hamiltonian}Speight-Ward Hamiltonian model}
\label{EnerCons}

The potential energy of the $\phi^4$ field [see equations (\ref{eq:lagrangy}) and (\ref{eq:Phi4Potential})] can be written as
\begin{eqnarray} \label{eq:PotEn}
E=\int_{-\infty}^{+\infty}\Big[\frac{1}{2}\phi_x^2+\frac{1}{2}(1-\phi^2)^2\Big]dx = \nonumber \\
\int_{-\infty}^{+\infty}\frac{1}{2}[\phi_x-(1-\phi^2)]^2dx+\int_{-\infty}^{+\infty} (1-\phi^2)\phi_x dx \equiv \tilde{E}+\frac{4}{3},
\end{eqnarray}
where the second integral was calculated as follows $\int_{-1}^{1} (1-\phi^2)d\phi=4/3$, taking into account the kink asymptotics $\phi \rightarrow -1$ as $x \rightarrow -\infty$ and $\phi \rightarrow 1$ as $x \rightarrow \infty$. Since the first integral has non-negative integrand, the minimal possible value of the potential energy is $E=4/3$ and this happens when
\begin{eqnarray} \label{eq:TanhKink}
\phi_x=1-\phi^2 \,\, \Rightarrow \,\,  \phi= \tanh(x-x_0).
\end{eqnarray}
This fact that the minimal energy for the solution satisfying the kink asymptotics is actually the kink solution is often referred to as a \jcmindex{\myidxeffect{B}!Bogomol'nyi bound} Bogomol’nyi bound~\cite{Bogomolnyi}.

\begin{figure}
\begin{center}
\includegraphics[width=7cm]{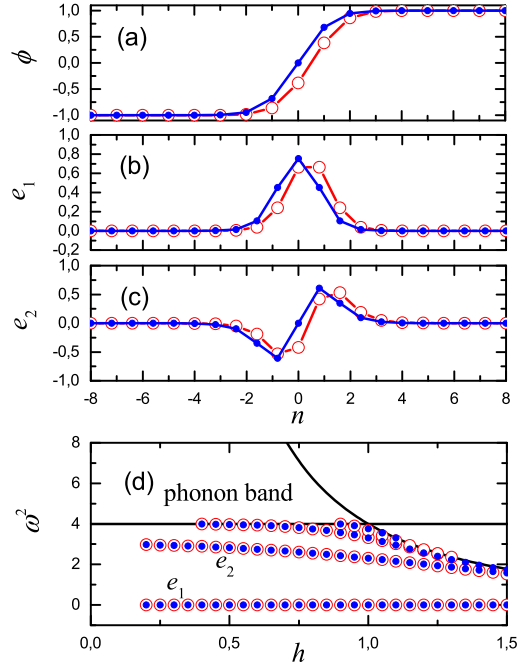}
\end{center}
\caption{Model (\ref{eq:SW}): (a) Kink profile, (b) Goldstone mode, and (c) kink vibrational mode for $h=0.8$. (d) Frequencies of the kink's internal modes for different magnitudes of the discreteness
parameter $h$. Two solid lines show the borders of the spectrum of vacuum, Eq. (\ref{SpecVacClassic1}). Results for the on-site and inter-site  kinks are shown by dots and circles, respectively. For any $h$ the kinks possess the zero frequency translational Goldstone mode in contrast to the kinks in standard model (\ref{eq:Classic}), see Fig.~\ref{fig1}. This means that in model (\ref{eq:SW}) the kinks are not trapped by the lattice and the PNp for them is precisely zero.}
\label{fig3}
\end{figure}

The idea of applying the Bogomol’nyi bound to the discrete case was proposed by Speight and Ward in a series of works \cite{SW,S1997,S1999}. They assumed that the on-site potential can be discretized using three neighboring nodes and considered a potential energy of the form
\begin{eqnarray} \label{E2}
   E= h \sum_n \left[ \frac 12 \left(\frac{\phi_{n+1}-\phi_n}{h}\right)^2 +\frac 12 F^2 \right],
\end{eqnarray}
where the on-site potential (second term) is taken in the form that one can write
\begin{eqnarray} \label{E2tr}
   E= h \sum_n  \frac 12 \left( \frac{\phi_{n+1}-\phi_n}{h} - F \right)^2 + \sum_n  (\phi_{n+1}-\phi_n)F.
\end{eqnarray}
If $F$ is chosen in a way that the second sum in (\ref{E2tr}) is telescopic, then the Bogomol’nyi bound can be achieved via evaluating the sum at $n \rightarrow \pm \infty$. For the $\phi^4$ model it was suggested to take $F=1 - (\phi_{n+1}^2 + \phi_{n+1}\phi_{n} + \phi_{n}^2)/3$ because for this choice
\begin{eqnarray} \label{DBound}
   \sum_n (\phi_{n+1}-\phi_n)F =\sum_n \left[ (\phi_{n+1}-\phi_n) -\frac 13(\phi_{n+1}^3-\phi_n^3) \right] =2-\frac 23 = \frac 43.
\end{eqnarray}
In this case again the minimum energy configuration is the kink (antikink) which can be found from the two-point map
\begin{equation}
u \equiv \pm\frac{\phi_{n+1}-\phi_{n}}{h}-1+\frac{\phi_{n}^2+\phi_{n}\phi_{n+1}+\phi_{n+1}^2}{3}=0.
\label{eq:phi4twopoint}
\end{equation}
The Hamiltonian of this exceptional discretization has the form:
\begin{equation}
H_2=\frac{h}{2} \sum\limits_{n}\left(\dot{\phi}_n^2+u^2 \right).
\label{eq:H2}
\end{equation}
The corresponding equation of motion is
\begin{eqnarray}\label{eq:SW}
\ddot{\phi}_n= \left(\frac{1}{h^2}+\frac{1}{3}\right)(\phi_{n-1}-2\phi_n+\phi_{n+1}) +2\phi_n \nonumber \\ -\frac{1}{9}\left[2\phi_n^3 +(\phi_n+\phi_{n-1})^3+(\phi_n+\phi_{n+1})^3\right].
\end{eqnarray}

The static kink solution can be found iteratively from the quadratic equation (\ref{eq:phi4twopoint})
\begin{eqnarray}
\phi_{n\pm1}=-\frac{\phi_n}{2}\mp\frac{3}{2h}\pm\frac{\sqrt{3}}{2}\sqrt{-\phi_n^2\pm\frac{6}{h}\phi_n+\frac{3}{h^2}+4},
\label{eq:Phi4kinksolution}
\end{eqnarray}
where one can take either upper or lower signs. The on-site kink can be obtained starting from the initial value $\phi_n=0$, while for the inter-site kink the initial value is $\phi_n=3/h-\sqrt{3+9/h^2}$ (given its placement around the origin, i.e., its anti-symmetry). A kink placed arbitrarily with respect to the lattice can be found starting from any initial value $|\phi_n|<1$, and all such kinks will have exactly same potential energy. This means that in the model (\ref{eq:SW}) static kinks do not experience PNp.

Linearization around an equilibrium solution, $\phi_n^0$, yields: 
\begin{eqnarray}
\ddot{\varepsilon}_n&=&\Delta_2\varepsilon_n+\frac{1}{3}\Big[1-(\phi_n^0+\phi_{n-1}^0)^2\Big]\varepsilon_{n-1}+\frac{1}{3}\Big[1-(\phi_n^0+\phi_{n+1}^0)^2\Big]\varepsilon_{n+1}\nonumber\\
&+&\frac{1}{3}\Big[4-2\phi_{n}^{0^{\scriptstyle 2}}-(\phi_n^0+\phi_{n-1}^0)^2-(\phi_n^0+\phi_{n+1}^0)^2\Big]\varepsilon_n.
\label{eq:Phi4SLiniriazation}
\end{eqnarray}
Substituting here the ansatz $\varepsilon_{n}=\exp(i k n -i
\omega t)$ one obtains the eigenvalue problem to find eigenfrequencies and eigenmodes of small-amplitude oscillations around the equilibrium solution $\phi_n^0$. For the spectrum of vacuum solution, $\phi_n^0=\pm1$, one finds 
\begin{eqnarray}
\omega^2=4+4\frac{1-h^2}{h^2}\sin^2\left(\frac{k}{2}\right).
\label{eq:Phi4SSpectrum}
\end{eqnarray}

In Fig.~\ref{fig3} we plot the same as in Fig.~\ref{fig1}, but for the model given by (\ref{eq:SW}), taking the same, relatively large value of the discreteness parameter, $h=0.8$. The results for the on-site and inter-site kinks are plotted by dots and circles, respectively. In (a) the kink profiles are shown. In (b) and (c) the two lowest frequency modes localized around the kinks are shown. The mode $e_1$ is the zero frequency Goldstone mode. The mode $e_2$, as before, represents the kink's internal vibration mode. In (d) solid lines give the lower and upper edges of the phonon band (\ref{eq:Phi4SSpectrum}) as a function of the discreteness parameter. The symbols represent the frequencies of the modes localized around the kink. The presence of the zero-frequency translational mode for any $h$ suggests that the kinks in model (\ref{eq:SW}) are not trapped by the lattice and that the PNp is precisely zero. 

\begin{figure}
\begin{center}
\includegraphics[width=8cm]{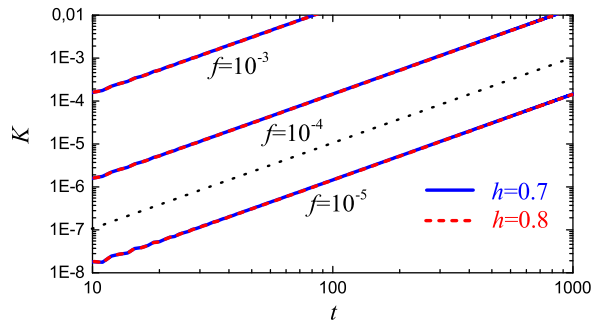}
\end{center}
\caption{Results for the Speight-Ward exceptional discretization (\ref{eq:SW}). Kink kinetic energy as a function of time for constant in time external force $f$, as specified for each curve. The discreteness parameter is $h=0.7$ (solid line) and $h=0.8$ (dashed line). The dotted line shows the slope equal to 2.}
\label{fig4}
\end{figure}

PNp-free kinks are accelerated by any weak external force. This is illustrated in Fig.~\ref{fig4}, which should be compared to Fig.~\ref{fig2}(b) plotted for the standard discrete model (\ref{eq:Classic}). Now we apply external force $f$ constant in time adding it to the right-hand side of (\ref{eq:SW}). The kink kinetic energy as a function of time is presented in log-log scale. The results  for $h=0.7$ (solid line) and $h=0.8$ (dashed line) practically coincide. The dotted line shows the slope equal to 2. It can be seen that $K\sim t^2$, which was expected for the uniformly accelerated kink motion with kink velocity $V_{\rm K}\sim t$ and kink kinetic energy $K\sim V_{\rm K}^2$. 

The PNp-free kinks can be boosted with the help of  the Goldstone mode. Suppose the equation of motion (\ref{eq:SW}) is integrated numerically with the time step $\Delta t$. The initial conditions can be taken as follows. At $t=0$ we set $\phi_n=\phi_n^0$, where $\phi_n^0$ is the static kink solution, and at $t=\Delta t$ the Goldstone mode is added, $\phi_n=\phi_n^0+s(e_1)_n$, with a small coefficient $s$ which defines the speed of the boosted kink. It is, of course, possible to excite simultaneously the translational Goldstone and kink's internal vibrational mode by using 
at $t=\Delta t$: 
$\phi_n=\phi_n^0+s(e_1)_n+v(e_2)_n$, with a small value of $v$, which defines the internal mode amplitude.

\subsection{Momentum conserving discretizations}
\label{MomCons}

Two different approaches to the derivation of momentum-conserving models will now be described.

\subsubsection{First approach}

In the previous Section it was shown, following the works \cite{SW,S1997,S1999}, that the telescopic summation was essential in the construction of the exceptional discretization. A different class of exceptional discretizations was offered in \cite{K2003} and again the telescopic summation was successfully used. This discretization is constructed not for the Hamiltonian but for the equation of motion. This approach leads to the model conserving momentum (\ref{mom4}), which is a discrete version of the \jcmindex{\myidxeffect{C}!Continuum momentum} continuum momentum $P=\int{\phi_t\phi_x dx}$; note that this expression
is often by convention used with a $(-)$ sign in front of the integral.

Conservation of momentum implies $dP/dt=0$. Differentiation of (\ref{mom4}) with respect to time gives
\begin{eqnarray} \label{dPdt}
\frac{dP}{dt} = \sum_n \left[ \ddot \phi_n (\phi_{n+1}-\phi_{n-1}) + \dot \phi_n (\dot \phi_{n+1}-\dot \phi_{n-1}) \right].
\end{eqnarray}
Note that due to the telescopic summation the second sum is zero. Let us see what happens when the standard  discretization of the $\phi^4$ field (\ref{eq:Classic}) is substituted in (\ref{dPdt}). One obtains
\begin{eqnarray} \label{dPdta}
\frac{dP}{dt} = \sum_n \Big[ \frac{1}{h^2}(\phi_{n+1}^2-\phi_{n-1}^2) - \left(\frac{2}{h^2}-2\right)(\phi_n \phi_{n+1}- \phi_n\phi_{n-1}) \nonumber \\
-2(\phi_n^3 \phi_{n+1}- \phi_n^3\phi_{n-1})\Big] =0.
\end{eqnarray}
The first two sums vanish due to their telescopic nature. For the first sum this becomes clear after rewriting it in the form $[(\phi_{n+1}^2+\phi_n^2)-(\phi_n^2+\phi_{n-1}^2)]/h^2$. However the third term does not disappear and we conclude that the standard discretization destroys the conservation of momentum. 

According to the work \cite{K2003}, the conservation of momentum can be achieved if the cubic term is discretized as $\phi^3 \rightarrow \phi_n^2(\phi_{n+1}+\phi_{n-1})/2$. The resulting exceptional discretization reads 
\begin{equation} \label{eq:MomCons}
\ddot{\phi}_n=\frac{1}{h^2}(\phi_{n-1}-2\phi_n+\phi_{n+1})+2\phi_n -\phi_n^2(\phi_{n+1}+\phi_{n-1}).
\end{equation}
The momentum is conserved because multiplying $\phi_n^2(\phi_{n+1}+\phi_{n-1})/2$ by $(\phi_{n+1}-\phi_{n-1})/2$ one gets $\phi_n^2(\phi_{n+1}^2-\phi_{n-1}^2)/4$, which is telescopic under summation and disappears, hence $dP/dt=0$.

The model (\ref{eq:MomCons}) conserves momentum (\ref{mom4}) but this is a non-Hamiltonian model and does not have a  conserved energy. It was shown that standard nearest-neighbour discretizations of Klein-Gordon models cannot conserve energy and linear momentum simultaneously \cite{DKY2006}.

\subsubsection{Second approach}

Another approach based on the methods of non-holonomic mechanics has been developed in \cite{KPR2015}. Let us consider the \jcmindex{\myidxeffect{L}!Lagrangian} Lagrangian of discretized Klein-Gordon equation in the form
\begin{eqnarray} \label{Lnonholonom}
   L= \sum_n \left[ \frac{1}{2} \dot{\phi}_n^2
   -\left( \frac{1}{2h^2} (\phi_{n+1}-\phi_n)^2
   +V(\phi_n) \right) \right].
\end{eqnarray}

In order to enforce the conservation law (\ref{mom4}) the following constraint in velocities is applied 
\begin{eqnarray} \label{Constr}
  \sum_n a_n(\boldsymbol \phi) \dot{\phi}_n = b(\boldsymbol \phi) \quad {\rm with} \quad a_n(\boldsymbol \phi) =\phi_{n+1}-\phi_{n-1}, \quad b(\boldsymbol \phi)=P,
\end{eqnarray}
where the notation $\boldsymbol \phi = (\phi_1, \phi_2, ... \phi_n)$ is introduced. This constraint, linear in velocity, is non-holonomic because it essentially depends on velocities and it cannot be integrated to contain only coordinates. The dynamics of the system under this type of constraints can be analyzed with the use of the Lagrange-d'Alembert principle~\cite{dAlambert1,dAlambert2}. For the constraints of the type (\ref{Constr}), the variations $\delta \phi_n$ are not arbitrary, but must satisfy the linear relations
\begin{eqnarray} \label{Relations}
  \sum_n a_n(\boldsymbol \phi) \delta \phi_n = \sum_n (\phi_{n+1}-\phi_{n-1})\delta \phi_n=0.
\end{eqnarray}

If $r$ is the Largange multiplier introduced to enforce the constraint (\ref{Relations}), the critical action principle respecting the constraint gives
\begin{eqnarray} \label{MinPrinciple}
  \int_{t_0}^{t_1}\left[ \frac{d}{dt} \frac{\partial L}{\partial \dot{\phi}_n} -\frac{1}{h^2}(\phi_{n+1}+\phi_{n-1}-\phi_n) +V^\prime(\phi_n) - r(\phi_{n+1}-\phi_{n-1}) \right]\delta \phi_n =0,
\end{eqnarray}
provided $\delta \phi_n$ vanish at $t = t_{0,1}$ so that no boundary integration terms appear in the equations. Taking into account that $\delta \phi_n$ is arbitrary, the equations of motion for the model conserving momentum are
\begin{equation} \label{eq:Model4}
\ddot{\phi}_n=\frac{1}{h^2}(\phi_{n-1}-2\phi_n+\phi_{n+1})-V^\prime (\phi_n) + r(\phi_{n+1}-\phi_{n-1}).
\end{equation}
Note the appearance of the new term proportional to $r$ on the right hand side of (\ref{eq:Model4}), as compared to (\ref{eq:Classic}). The physical meaning of $r$ is the constraint force that is chosen to ensure momentum conservation. One can find $r$ by introducing (\ref{eq:Model4}) into (\ref{dPdt}). The linear terms vanish telescopically and the nonlinear term disappears when
\begin{equation} \label{eq:r}
r=\frac{\sum_n V^\prime(\phi_n)(\phi_{n+1}-\phi_{n-1})} {\sum_n (\phi_{n+1}-\phi_{n-1})^2}.
\end{equation}
We need to demonstrate that the last term in (\ref{eq:Model4}) vanishes in the continuum limit $h \rightarrow 0$. The continuum analog of $r(\phi_{n+1}-\phi_{n-1})$ is $\phi_x(\int V^\prime (\phi)\phi_x dx) / \int \phi_x^2 dx$. The integral in the numerator can be integrated to give $V(\phi_2)-V(\phi_1)$ which is zero for the Klein-Gordon field supporting static kinks with the asymptotics $\phi_1$ for $x \rightarrow -\infty$ and $\phi_2$ for $x \rightarrow \infty$.

In the work \cite{KPR2015} kink mobility was analyzed for model (\ref{eq:Model4}) and for the standard discretization (\ref{eq:Classic}). It was shown that even in the regime of high discreteness, where the standard discretization produces kinks that are trapped by PNp, kinks in model (\ref{eq:Model4}) are mobile.

Another remark is that model (\ref{eq:Model4}) does not conserve energy. Multiplying both sides of (\ref{eq:Model4}) with $r$ defined in (\ref{eq:r}) by $\dot{\phi}_n$ and summing we find that instead of $dH/dt=0$, which was true for the Hamiltonian (\ref{H1}), now the following holds \cite{KPR2015}
\begin{equation} \label{eq:Hnew}
\frac{dH}{dt}=rP.
\end{equation}

\subsection{Discretized first integral approach}
\label{Sec:DFI}

We saw that the static $\phi^4$ kink solutions for the exceptional discretizations derived in Sec.~\ref{EnerCons} and Sec.~\ref{MomCons} can be found from a two-point map starting from any initial value $-1<\phi_n<1$. With the use of the \jcmindex{\myidxeffect{D}!Discretized first integral} discretized first integral (DFI) approach \cite{journal2} we will obtain the map from which not only kinks but a one-parameter family of static solutions can be obtained.

We consider the static Klein-Gordon equation (\ref{eq:staticform}) and denote
\begin{equation} \label{D}
   D(\phi(x)) \equiv \phi_{xx} -V^\prime(\phi)=0\,.
\end{equation}
The first integral of (\ref{D}) is
\begin{equation} \label{FI}
   U(x) \equiv \phi_{x}^2 -2V(\phi)+C=0\,,
\end{equation}
where $C$ is the integration constant. The first integral can also be taken in a modified form, e.g., as
\begin{equation} \label{FIa}
   u(x) \equiv \pm \phi_{x} -\sqrt{2V(\phi)-C}=0\,.
\end{equation}

We will consider discrete versions of (\ref{FI}) and (\ref{FIa}) in the form
\begin{equation} \label{DFI}
   U(h,\phi_{n-1},\phi_n) \equiv \frac{(\phi_{n}-\phi_{n-1})^2}{h^2} -2V(\phi_{n-1},\phi_{n})+C=0\,,
\end{equation}
and
\begin{equation} \label{DFIa}
   u(h,\phi_{n-1},\phi_n) \equiv \pm \frac{\phi_{n}-\phi_{n-1}}{h} -\sqrt{2V(\phi_{n-1},\phi_{n})-C}=0\,,
\end{equation}
respectively, where we demand that $V(\phi_{n-1},\phi_{n})\rightarrow V(\phi)$  when $h \rightarrow 0$.

\subsubsection{Momentum conserving model}

Considering the first integral of the form (\ref{FI}) we calculate $dU/dx$ and multiply the result by $(dx/d\phi)/2$ to obtain
\begin{equation} \label{DU}
   \frac 12 \frac{dU}{d\phi}=D\,.
\end{equation}
Substituting the left-hand side of (\ref{DU}) with its discrete version and restoring the dynamical term we obtain the discrete model after \cite{K2003} which conserves momentum (\ref{mom4}):
\begin{equation} \label{MomConsAgain}
   \ddot{\phi}_n= \frac{U(h,\phi_{n},\phi_{n+1})-U(h,\phi_{n-1},\phi_{n})}{\phi_{n+1}-\phi_{n-1}}\,.
\end{equation}
Obviously, static solutions of this model can be found from the two-point map (\ref{DFI}) which contains a free parameter $C$.  

Particularly, considering the discrete version of the $\phi^4$ potential (\ref{eq:Phi4Potential}) in the form 
\begin{equation} \label{Vphi4Discr}
   V(\phi_{n-1},\phi_{n})=\frac 12 (1-\phi_{n-1}\phi_{n})^2\,,
\end{equation}
from (\ref{DFI}) and (\ref{MomConsAgain}) we get the equation of motion of model (\ref{eq:MomCons}) which does not include the integration constant $C$. All static solutions of model (\ref{eq:MomCons}) can be found from the two-point map defined by (\ref{DFI}) and (\ref{Vphi4Discr}). This map is a quadratic equation with the roots: 
\begin{equation} \label{QuadrSol}
   \phi_n=\frac {(1-h^2)\phi_{n-1}\pm \sqrt{\mathcal{D}}/2}{1-h^2\phi_{n-1}^2}, \quad \mathcal{D}=4h^2(1-\phi_{n-1}^2)^2 +4h^2(h^2\phi_{n-1}^2-1)C,
\end{equation}
where, due to the symmetry of the equation, one can interchange $\phi_n$ and $\phi_{n-1}$.

For any chosen value of the integration constant $C$, starting from any initial value of $\phi_n$ for which $\mathcal{D}\ge 0$ and $h^2\phi_{n-1}^2 \neq 1$, one can obtain from (\ref{QuadrSol}) a static solution of model (\ref{eq:MomCons}). The kink solution is found for $C=0$ and any $-1<\phi_n<1$. Solutions for other values of $C$ were analyzed in \cite{journal2}. It was shown that those solutions can be expressed in terms of the Jacobi elliptic functions.

\subsubsection{Energy conserving models}

We now modify (\ref{MomConsAgain}) as follows
\begin{equation} \label{MomConsAgainModif}
   \ddot{\phi}_n= e(h,\phi_n)\frac{U(h,\phi_{n},\phi_{n+1})-U(h,\phi_{n-1},\phi_{n})}{\phi_{n+1}-\phi_{n-1}}\,,
\end{equation}
assuming that $e(h,\phi_n)$ is a continuous function that 
never vanishes and $e(h,\phi_n)\rightarrow 1$ in the continuum limit ($h \rightarrow 0$). If so, this multiplier will not affect the static solutions of this exceptional discretization (\ref{MomConsAgain}) and will not change its continuum limit. For the particular choice $e(h,\phi_n)=1/(1-h^2\phi_n^2)$ one can obtain from (\ref{MomConsAgainModif}), (\ref{DFI}) and (\ref{Vphi4Discr}) 
the following model discovered in \cite{CKMS2005}
\begin{equation} \label{EnConsAgain}
   \ddot{\phi}_n= \frac{1}{h^2}(\phi_{n-1}-2\phi_{n}+\phi_{n+1}) + \frac{2(\phi_{n}-\phi_{n}^3)}{1-h^2\phi_{n}^2}\,,
\end{equation}
which does not depend on $C$ and possesses the Hamiltonian 
\begin{equation}
H_4=\frac{h}{2} \sum\limits_{n}\left[\dot{\phi}_n^2+ \frac{(\phi_n-\phi_{n-1})^2}{h^2}+ V(\phi_n) \right],
\label{eq:H4}
\end{equation}
with the potential
\begin{equation}
V(\phi_n) =-\frac{1}{h^2} \left(\phi_n^2+ \frac{1-h^2}{h^2} \ln \left| 1- h^2\phi_n^2 \right|  \right).
\label{eq:M4potential}
\end{equation}
This energy-conserving exceptional discretization is of particular interest because here the on-site potential is discretized locally. Nevertheless, so far there does not appear to exist a general approach to the derivation of except ional discretizations having this property. 

Static solutions of model (\ref{eq:Model4}) coincide with those of momentum-conserving model (\ref{eq:MomCons}) and can be found iteratively with the help of (\ref{QuadrSol}). On the other hand, dynamic properties of models (\ref{eq:MomCons}) and (\ref{EnConsAgain}) are different. Stability of static solutions of model (\ref{eq:MomCons}) and model (\ref{eq:Model4}) was analyzed in \cite{journal2}. The kink solution is stable in both models, yet other solutions can, in principle, be stable in one model and unstable in the other.

The $\phi^4$ models (\ref{eq:MomCons}) and (\ref{EnConsAgain}) do not include the integration constant $C$ even though their DFI does. Generally speaking, the DFI approach produces discrete models that include $C$. Let us demonstrate this for the Speight-Ward model \cite{SW,S1997,S1999}. We rewrite the potential energy of the Hamiltonian (\ref{eq:lagrangy}) with the use of (\ref{FIa}) in the following form
\begin{eqnarray} \label{eq:PotSW}
E=\int_{-\infty}^{+\infty}\Big[[u(x)]^2+2\phi_x \sqrt{2V(\phi)-C}\Big]dx,
\end{eqnarray}
where the constant term was omitted. Discretizing this potential energy and adding the kinetic energy term we obtain the Hamiltonian 
\begin{equation} \label{eq:H4SW}
H=\frac{h}{2} \sum\limits_{n}\left[\dot{\phi}_n^2+ [u(h,\phi_{n-1}, \phi_n)]^2 +2\frac{\phi_n-\phi_{n-1}}{h} \sqrt{2 V(\phi_{n-1}, \phi_n)-C} \right].
\end{equation}
The on-site potential is discretized following the works \cite{SW,S1997,S1999} in the form
\begin{eqnarray} \label{eq:SWG}
\sqrt{2 V(\phi_{n-1}, \phi_n)-C}= \frac{G(\phi_n)-G(\phi_{n-1})}{\phi_{n}-\phi_{n-1}}, \nonumber \\
{\rm where} \quad G^\prime(\phi)=\sqrt{2 V(\phi)-C}.
\end{eqnarray}
With this choice the last term of (\ref{eq:H4SW}) disappears in telescopic summation. The exceptional discretization will have the Hamiltonian
\begin{equation} \label{eq:H4SWtild}
\tilde{H}=\frac{h}{2} \sum\limits_{n}\left[\dot{\phi}_n^2+ [\tilde{u}(h,\phi_{n-1}, \phi_n)]^2  \right],
\end{equation}
where according to (\ref{DFIa}) and (\ref{eq:SWG}) 
\begin{equation} \label{eq:H4SWtild1}
\tilde{u}(h,\phi_{n-1}, \phi_n)=\frac{\phi_n-\phi_{n-1}}{h}-\frac{G(\phi_n)-G(\phi_{n-1})}{\phi_{n}-\phi_{n-1}}.
\end{equation}
This model and DFI $\tilde{u}(h,\phi_{n-1}, \phi_n)=0$ include the integration constant $C$ through the function $G$.

\subsubsection{More examples of exceptional discrete $\phi^4$ models}

The following intriguing PNp-free model was derived by Barashenkov {\it et al.} \cite{BOP2005}
\begin{equation} \label{Barash}
   \ddot{\phi}_n= \frac{1}{h^2}(\phi_{n-1}-2\phi_{n}+\phi_{n+1}) + 2\phi_{n}-2\phi_{n-1}\phi_{n}\phi_{n+1}\,,
\end{equation}
and it was subsequently found \cite{DKKS2007} that it conserves momentum in the form 
\begin{eqnarray} \label{mom4Kev}
P=\sum_n \dot{\phi}_n (\phi_{n+2}-\phi_{n-2}),
\end{eqnarray}
which is different from (\ref{mom4}).

All static solutions of (\ref{Barash}) can be found from the following DFI \cite{DKKS2007} 
\begin{eqnarray} \label{QuadrSolX}
   \phi_{n\pm 1}=(2-\Lambda)\frac {Z\phi_{n}\pm \sqrt{\mathcal{D}}}{2-\Lambda-\Lambda \phi_{n}^2},
\end{eqnarray}
where
\begin{eqnarray} \label{QuadrSolParam}
    \mathcal{D}=\frac{\Lambda}{2-\Lambda} (1-X\phi_n^2+\phi_n^4-C), \quad \Lambda=2h^2, \nonumber \\
   Z=\frac{1-2h^2+Ch^4}{1-h^2}, \quad X=\frac{(1-C)\Lambda^2+ (2-\Lambda)^2(1-Z^2)}{\Lambda(2-\Lambda)},
\end{eqnarray}
and $C$ is the integration constant. One can substitute $\phi_{n\pm 1}$ with $\phi_{n\mp 1}$ in the left-hand side of (\ref{QuadrSolX}) due to the symmetry of the equation. For any pair of $C$ and admissible initial value $\phi_0$ the two-point map (\ref{QuadrSolX}) produces a static solution to (\ref{Barash}). The map has two different roots and for $\phi_{n+1}$ (or for $\phi_{n-1}$) one must take the one different from $\phi_{n-1}$ (or from $\phi_{n+1}$). Analysis of the static solutions to (\ref{Barash}) and their expression in terms of the Jacobi elliptic functions can be found in \cite{DKKS2007}.



\subsection{A unifying view}
\label{Sec:Unif}

Several different approaches to derivation of exceptional discretizations have been described above, including ones conserving energy or momentum. Such models have different physical properties, but there should be something in common between them. As suggested in \cite{BOP2005}, all exceptional discretizations must have static second order (three-point) difference problem reducible to a first-order (two-point) map. Then a static solution $\phi_n=g(nh-x_0)$ which exists for an arbitrary central position $x_0$ can be derived iteratively. 
For the monotonically growing function $g$ describing a kink solution this can be inverted to obtain $nh-x_0=g^{-1}(\phi_n)$ so that
\begin{eqnarray} \label{Invert}
   \phi_{n+1}=g(g^{-1}(\phi_n)+h).
\end{eqnarray}
This means that for a translationally invariant static solution a two-point map can be defined. Then the following strategy for the construction of exceptional discretizations has been offered \cite{BOP2005}, taking a two-point map in the form 
\begin{eqnarray} \label{MapGen}
   \frac{\phi_{n+1}-\phi_n}{h}=F(\phi_{n+1},\phi_n),
\end{eqnarray}
where for the $\phi^4$ case (\ref{eq:Phi4Potential}) it is assumed that $F(\phi_{n+1},\phi_n) \rightarrow 1-\phi^2$ when $h \rightarrow 0$, to ensure correct continuum limit
\begin{eqnarray} \label{ContDifference}
   \phi_x=1-\phi^2.
\end{eqnarray}
The static kink solution (\ref{eq:Kink}) with $v=0$ to (\ref{eq:staticform}) is also a static solution to its first integral (\ref{ContDifference}). Now we square both sides of (\ref{MapGen}) and of the shift $(\phi_{n}-\phi_{n-1})/h=F(\phi_{n},\phi_{n-1})$ and find their difference 
\begin{eqnarray} \label{Difference}
   \frac{\phi_{n-1}-2\phi_n+\phi_{n+1}}{h^2}=\frac{F^2(\phi_{n+1},\phi_n)-F^2(\phi_{n},\phi_{n-1})}{\phi_{n+1}-\phi_{n-1}}.
\end{eqnarray}
This is an exceptional discretization and it is clear now that it has a common property with the energy conserving model (\ref{eq:SW}) and momentum conserving model (\ref{eq:MomCons}), namely, static translationally invariant solutions for all exceptional discretizations are derivable from two-point maps.

\section{Conserved quantities for exceptional discretizations of the $\phi^4$ field}
\jcmindex{\myidxeffect{C}!Conservation law}

In \cite{BOP2005,KDS2009} a generalized discretization of the $\phi^4$ field was analyzed in the form
\begin{eqnarray} \label{phi4General}
   \ddot{\phi}_n= \frac{1}{h^2}(\phi_{n-1}-2\phi_{n}+\phi_{n+1}) + 2\phi_{n} - A_1\phi_n^3 -\frac{A_2}{2} \phi_n^2(\phi_{n+1}+\phi_{n-1}) \nonumber \\
   -\frac{A_3}{2} \phi_n(\phi_{n+1}^2+\phi_{n-1}^2) -A_4\phi_{n-1}\phi_{n}\phi_{n+1} \nonumber \\ 
   -\frac{A_5}{2}\phi_{n+1}\phi_{n-1}(\phi_{n+1}+\phi_{n-1})-\frac{A_6}{2}(\phi_{n+1}^3+\phi_{n-1}^3)\,,
\end{eqnarray}
where model parameters satisfy
\begin{eqnarray} \label{AkGeneral}
   \sum_{k=1}^{6}A_k=2\,,
\end{eqnarray}
to ensure the correct continuum limit.

In this general form the model is not an exceptional discretization, but it includes the above mentioned exceptional discretizations as special cases, except for the Hamiltonian model (\ref{EnConsAgain}), which is not a member of (\ref{phi4General}). 

It can be straightforwardly verified that if
\begin{eqnarray} \label{MconsGen}
   A_1=\frac{A_3}{2}=A_4=2\delta, \quad A_2=2(1-4\gamma-4\delta), \quad A_5=A_6=4\gamma ,
\end{eqnarray}
for arbitrary $\delta$ and $\gamma$, then the model (\ref{phi4General}) is a special discretization conserving momentum (\ref{mom4}). All static solutions of this model can be found from the two-point map
\begin{eqnarray} \label{MapMconsGen}
   U(\phi_{n-1},\phi_n)=\frac{(\phi_{n}-\phi_{n-1})^2}{h^2} +2\phi_{n}\phi_{n-1} -2\gamma(\phi_{n}^4+\phi_{n-1}^4) \nonumber \\ 
-2\delta\phi_{n}\phi_{n-1}(\phi_{n}^2+\phi_{n-1}^2) +2(2\gamma+2\delta-\frac 12)\phi_{n}^2\phi_{n-1}^2 -C=0 ,
\end{eqnarray}
where $C$ is the integration constant. This is so because in this case (\ref{phi4General}) can be written in the form (\ref{MomConsAgain}). The two-point map (\ref{MapMconsGen}) is DFI, i.e., in the continuum limit it reduces to (\ref{FI}). Note that for $\delta=\gamma=0$ (\ref{mom4Kev}) is the so-called 
\jcmindex{\myidxeffect{B}!Bender-Tovbis model} Bender-Tovbis model, see \cite{BT1997}. 

As was already stated before, momentum (\ref{mom4Kev}) is conserved by (\ref{phi4General}) only if $A_4=2$ and all other $A_k$ are zero \cite{DKKS2007}.


Model (\ref{phi4General}) with parameters satisfying
\begin{eqnarray} \label{HconsGen}
   A_1=8\alpha_1, \quad A_2=12\alpha_2, \quad A_3=8\alpha_3, \quad A_4=A_5=0, \quad A_6=4\alpha_2 , \nonumber \\
   {\rm with} \quad \alpha_1+2\alpha_2+\alpha_3=1/4,
\end{eqnarray}
conserves the total energy
\begin{equation} \label{eq:H4Gen}
H=h\sum\limits_{n}\left[\frac{\dot{\phi}_n^2}{2}+\frac{(\phi_n-\phi_{n-1})^2}{2h^2} - \phi_{n}^2 +\alpha_1\phi_{n}^4 +\alpha_2\phi_{n}\phi_{n-1}(\phi_{n}^2+\phi_{n-1}^2) +\alpha_3\phi_{n}^2\phi_{n-1}^2 \right].
\end{equation}
This model has two free parameters. The Hamiltonian Speight-Ward model (\ref{eq:SW}) \cite{SW,S1997,S1999} corresponds to $\alpha_1=\alpha_2=1/18$ and $\alpha_3=1/12$.

\section{Exact moving kink solutions for the discrete $\phi^4$ models}
\jcmindex{\myidxeffect{M}!Moving kink solution}

In the work \cite{journal1} exact moving kink solutions for the $\phi^4$ discretization (\ref{phi4General}), (\ref{AkGeneral}) have been derived. The solution of the form
\begin{equation} \label{eq:MovKinkGen}
\phi_n(x,t)=\tanh[\beta(hn+hx_0-vt)],
\end{equation}
where $v$ is the propagation velocity, $\beta$ is the inverse width, and $x_0$ is the initial position of the kink, exists under the following constraints
\begin{equation} \label{eq:C67}
\beta^2=-\frac{A_1}{2v^2},
\end{equation}
\begin{equation} \label{eq:C68}
2A_3=-A_5(1+T),
\end{equation}
\begin{equation} \label{eq:C69}
A_6=0,
\end{equation}
\begin{equation} \label{eq:C70}
\frac{2T}{h^2}=A_2+A_4(1+T)+\frac{A_5}{2}(1+2T-T^2),
\end{equation}
\begin{equation} \label{eq:C71}
\frac{2(1-h^2)T}{h^2}=-A_1T+A_2(1-T)+A_4+\frac{A_5}{2}(1+T),
\end{equation}
where
\begin{equation} \label{eq:C72}
T=\tanh^2(h\beta).
\end{equation}

Interestingly, this exact kink solution can have an arbitrarily large velocity, though this is possible only for non-Hamiltonian models. To make this more transparent a particular case will be considered when only $A_1$, $A_2$, and $A_4$ are nonzero. Taking into account the continuum limit constraint (\ref{AkGeneral}), this model has a free parameter $A_1<0$. From (\ref{eq:C67})-(\ref{eq:C71}) and (\ref{AkGeneral}) one has
\begin{equation} \label{eq:C73}
\beta^2=-\frac{A_1}{2v^2}, \quad T=\tanh^2(h\beta), \quad A_4=\frac{2}{h^2}- \frac{2-A_1}{T}, \quad A_2=2-A_1-A_4.
\end{equation}
It is clear that, for any $A_1<0$, one can take any value of $v$ and then find from (\ref{eq:C73}) parameters $\beta$, $A_4$, and $A_2$. From the first relation in (\ref{eq:C73}) it is clear that the kink becomes wider for larger velocity, which is opposite to the continuum $\phi^4$ kink (\ref{eq:Kink}). 

Interestingly, the kink (\ref{eq:MovKinkGen}) with the parameters (\ref{eq:C73}) can have arbitrary large velocity for any small but finite value of $h$. Taking into account that for small $h$ we have $T\approx h^2\beta^2$, the third expression in (\ref{eq:C73}) assumes the form $A_4=[2-(2-A_1)/\beta^2]/h^2$. From here it is clear that for $h \rightarrow 0$ the coefficients $A_2$ and $A_4$ diverge having opposite signs.

Recall that the model (\ref{phi4General}) is Hamiltonian when (\ref{HconsGen}) is satisfied. From the analysis of (\ref{eq:C67})-(\ref{eq:C71}) it is not difficult to come to the conclusion that among the models supporting moving kinks there are no Hamiltonian models.



\section{Collision of translationally invariant $\phi^4$ kinks}
\label{Sec:Collisions}
\jcmindex{\myidxeffect{K}!Kink collision}

\begin{figure}
\begin{center}
\includegraphics[width=8cm]{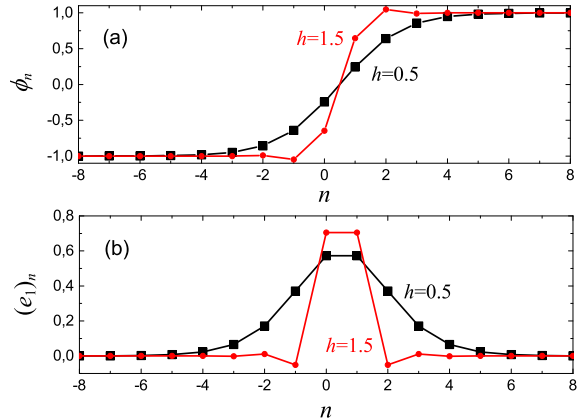}
\end{center}
\caption{Results for the model of Eq.~(\ref{eq:SW}). (a) Inter-site kink profile and (b) Goldstone mode for $h=0.5$ (squares) and $h=1.5$ (dots). For $h<1$ kink tails are smooth but for $h>1$ they oscillate near the asymptotic values $\pm 1$.}
\label{fig5}
\end{figure}

\begin{figure}
\begin{center}
\includegraphics[width=12cm]{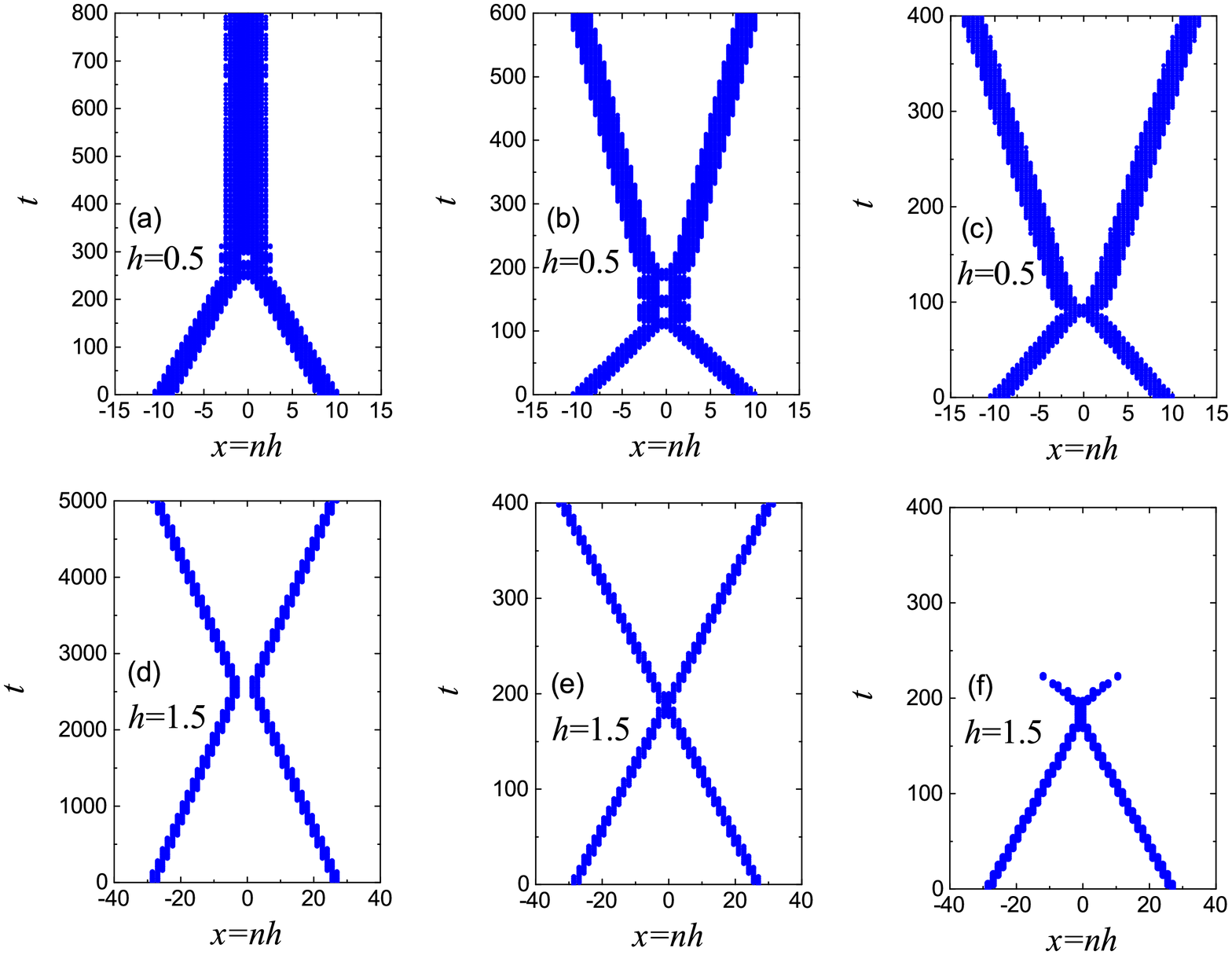}
\end{center}
\caption{Results for the model of Eq.~(\ref{eq:SW}). Kink-antikink collisions for (a-c) $h=0.5$ and (d-f) $h=1.5$. Shown are particles in the $(x,t)$ plane having total energy greater than $0.2E_{\max}$, where $E_{\max}$ is the maximal energy. Kink velocity before the collision is: (a) 0.03, (b) 0.073, (c) 0.09, (d) 0.01, (e) 0.15, and (f) 0.155. In (a) kink-antikink annihilation with formation of a standing bion takes place. In (f) kink-antikink annihilation produces rapidly spreading radiation. Kink velocity after collision is (b) 0.018, (c) 0.035, (d) 0.01, (e) 0.15.}
\label{fig6}
\end{figure}

Enhanced kink mobility in the models free of static PNp makes it possible to study kink collisions even in the regime of high discreteness, which is impossible in the standard discretization, given the deceleration and eventual pinning of the kinks in the latter due to the PNp. The translational Goldstone mode can be used for boosting the kinks, as  was described in the last paragraph of Sec.~\ref{EnerCons}. 


Kink collisions in different discrete $\phi^4$ models were analyzed in \cite{Roy2007} for $h\leq 0.3$, i.e. close to the continuum limit. Collisions in the standard discrete model of 
Eq.~(\ref{eq:Classic}) were compared to those in the PNp-free energy conserving models (\ref{eq:SW}) and (\ref{EnConsAgain}) and in two momentum conserving models. Overall it was concluded that kink collisions in the translationally invariant models are more elastic as compared to the standard discrete model, because the threshold escape velocity in the PNp-free models is always smaller. 

To the best of our knowledge, kink-antikink collisions in the regime of high discreteness have not been analyzed because for $h\sim 1$ the PNp is typically so high that kinks can hardly propagate. Our aim here is to demonstrate the possibility of such studies for the exceptional discretizations, rather than to do a comprehensive analysis even for one particular model. 

We choose the Hamiltonian model of Eq.~(\ref{eq:SW}) 
and compare kink-antikink collisions for $h=0.5$ and $h=1.5$. From Fig.~\ref{fig3}(d) one can see that the lines showing the borders of the phonon spectrum cross each other at $h=1$. That is why, for $h<1$ kinks have smooth tails, while for $h>1$ their tails oscillate near the asymptotic values $\pm 1$, see Fig.~\ref{fig5}(a). 
Recall that the exact static kink solutions can be found from  the two-point map (\ref{eq:Phi4kinksolution}). In Fig.~\ref{fig5}(b) the Goldstone translational mode is shown for the kinks presented in (a). This mode is found by solving the eigenvalue problem for the equation of motion (\ref{eq:Phi4SLiniriazation}) linearized in the vicinity of static kink solution. The results for $h=0.5$ are shown by squares and for $h=1.5$ by circles. 

The Goldstone mode is then used for kink boosting, with the moving kink colliding with its mirror image antikink. Collision results are presented in Fig.~\ref{fig6}, where particles having total energy greater than $0.2E_{\max}$, where $E_{\max}$ is the maximal energy, are shown in the $(x,t)$ plane. This way the trajectories of the colliding kinks and antikinks are visualized. 
Let us first discuss the results for relatively small discreteness, $h=0.5$, shown in the top panels of Fig.~\ref{fig6}. The collision velocity increases from the left to the right. In (a) kink and antikink collide with a small velocity of 0.03, which is below the escape threshold value. A bound state arises in the form of a bion, which gradually radiates its energy. In (b) the collision velocity is 0.073, which is close to the escape threshold. A three-bounce collision 
ends with kink separation with final velocity $\approx 0.018$. This resonant effect in kink-antikink collisions is due to the energy exchange between kink's translational and internal vibration modes, as described in a number of studies for the continuum $\phi^4$ equation \cite{Campbell,anninos,Campbell1,goodman} (and also in multiple other Chapters in this Special Volume). In (c) the collision velocity is 0.09, which is above the threshold escape velocity. This inelastic collision is accompanied by radiation of phonon wave packets and the final kink velocity reduces to 0.035.

The results for the case of high discreteness, $h=1.5$, are shown in the bottom panels of Fig.~\ref{fig6}. Here, as well, the collision velocity increases from the left to the right. It turns out that the kink and antikink having oscillatory tails are repulsive solitons in contrast to the mutually attractive kink and antikink with smooth tails. This is the reason of qualitatively different picture of collisions observed for large discreteness. 
In (d) the collision velocity is 0.01, which is not sufficient to overcome the apparent repulsion and the cores of the kink and antikink do not merge. As a result, the collision appears to be approximately elastic. In (e) the collision velocity is 0.15, which is close but still below the value needed to overcome mutual repulsion of the kink and antikink, and again the collision is practically elastic. Finally in (f), the collision velocity of 0.155 is sufficient to overcome the repulsion between colliding solitons and this collision is strongly inelastic resulting in a fast decay of the solitons into small-amplitude phonon wave packets.

It can be concluded that the PNp-free discrete models give the possibility to study collisions between kinks and antikinks in the regime of high discreteness and new effects can be observed in this regime,
including the apparent kink repulsion and the potential kink destruction as a result of their collisions.
Nevertheless, these preliminary observations clearly warrant further and more systematic
study.

\section{Conclusions \& Future Challenges}
In the present review, we have revisited the different methods that can produce
exceptional discretizations in $\phi^4$ models, as a special case of more general
Klein-Gordon type chains. We have illustrated the undesirable features of standard
discretizations, such as the exponentially increasing (with the lattice spacing $h$)
Peierls-Nabarro potential and the resulting deceleration, eventual trapping and 
pinning of kinks. A side product of this is the inability to consider kink-antikink
collisions and their potential energy-exchange mechanisms. We saw that these
features (including also the finite external dc force needed to de-pin the kinks in the
above standard case)  typically disappear in exceptional discretizations and provided
a diverse array of methods that can produce such discretizations. These consisted of energy-conserving
ones (including the ones produced in the work of Speight (and collaborators)~\cite{SW,S1997,S1999}
but also others such as~\cite{CKMS2005}), as well as momentum conserving ones (including
some of the work of the present authors); we also illustrated the connections between the
two and the relevance of the discretized first integral approach. The resulting discretizations
not only featured the presence of arbitrarily centered kinks and a neutral (Goldstone-like)
eigendirection. They also enabled the consideration of collisions between kinks and antikinks
even in the highly discrete realm.

At the same time, this discussion paved the way for a number of quite important
questions. While the case of weak discreteness has seen some analysis at the level
of kink-antikink interactions and how they compare for different models, the territory
is far less charted in the case of strong discreteness and large values of $h$. 
We have only scratched the tip of the iceberg in that regard illustrating unusual
features in the case of discretizations of the type of~\cite{S1997}. These included
the non-exponential (spatial) decay of the kink for $h>1$ in this model, as well as the 
detrimental effect of interactions for suitably large velocities. It would be particularly
interesting to further explore these features and whether they may arise in other exceptional
discretization models as well.

Most of the models considered herein have been motivated by the (in some ways, favorable) properties
of the respective discretizations. However, 
there also exist discrete physical models supporting PNp-free kinks, e.g., a chain of electric dipoles rotating in a plane containing the chain \cite{SZ2006} or kinks in topological mechanical chains \cite{MechChain2017}. The search for other applications of such discrete systems is of great importance and the more detailed consideration of the properties of such physically relevant models
is of paramount importance in its own right.

\section*{Acknowledgments}

The work of S.V.D. was supported by the grant of the Russian Science Foundation (No. 16-12-10175).
This material is based upon work supported 510
by the National Science Foundation under Grant No. DMS-1809074 (P.G.K).

%


\end{document}